\documentclass[conference]{IEEEtran}
\IEEEoverridecommandlockouts
\IEEEpubid{\makebox[\columnwidth]{978-3-903176-15-7~\copyright~2019 IFIP \hfill} \hspace{\columnsep}\makebox[\columnwidth]{ }}
\usepackage{cite}
\usepackage{amsmath,amssymb,amsfonts}
\usepackage{algorithmic,algorithm}
\usepackage{graphicx}
\usepackage{subfig}
\usepackage{textcomp}
\usepackage{xcolor}
\usepackage{color}
\def\BibTeX{{\rm B\kern-.05em{\sc i\kern-.025em b}\kern-.08em
    T\kern-.1667em\lower.7ex\hbox{E}\kern-.125emX}}
\begin{document}

\title{Optimizing Controller Placement for Software-Defined Networks}

\author{\IEEEauthorblockN{Victoria Huang, Gang Chen, Qiang Fu}
\IEEEauthorblockA{\textit{School of Engineering and Computer Science} \\
\textit{Victoria University of Wellington}\\
Wellington, New Zealand \\
guiying.huang, aaron.chen, qiang.fu@ecs.vuw.ac.nz}
\and
\IEEEauthorblockN{Elliott Wen}
\IEEEauthorblockA{\textit{School of Computer Science} \\
\textit{The University of Auckland}\\
Auckland, New Zealand \\
jwen929@aucklanduni.ac.nz}
}

\maketitle

\begin{abstract}
Controller placement problem (CPP) is a key issue for Software-Defined Networking (SDN) with distributed controller architectures. This problem aims to determine a suitable number of controllers deployed in important locations so as to optimize the overall network performance. In comparison to communication delay, existing literature on the CPP assumes that the influence of controller workload distribution on network performance is negligible.  

In this paper, we tackle the CPP that simultaneously considers the communication delay, the control plane utilization, and the controller workload distribution. Due to this reason, our CPP is intrinsically different from and clearly more difficult than any previously studied CPPs that are NP-hard. To tackle this challenging issue, we develop a new algorithm that seamlessly integrates the genetic algorithm (GA) and the gradient descent (GD) optimization method. Particularly, GA is used to search for suitable CPP solutions. The quality of each solution is further evaluated through GD. Simulation results on two representative network scenarios (small-scale and large-scale) show that our algorithm can effectively strike the trade-off between the control plane utilization and the network response time.

\end{abstract}


\section{Introduction}
Software-Defined Networking (SDN) is an emerging networking paradigm notable for decoupling the network control from the data plane and forming a logically centralized external control plane. It enables centralized network management and significantly speeds up network innovation. Traditionally, the control plane is equipped with one single controller. As the network scales up, architectures supporting distributed controllers (e.g., ONOS \cite{onos}) have been successfully developed to substantially increase the combined processing capacity of the control plane in order to handle the growing demand for traffic processing.

The increasing popularity of distributed controller architectures also gives rises to new research problems among which an essential one is the \textit{Controller Placement Problem} (CPP) \cite{controllerPlacement,yao2014capacitated,hu2013reliability}. The CPP aims to identify a suitable number of controllers as well as their locations so as to optimize the network performance. For example, the CPP is first introduced by \textit{Heller et al.} \cite{controllerPlacement} in an attempt to optimize controller placement by minimizing the communication delay among switches and controllers. Recently, several approaches \cite{hu2017reliable, hu2013reliability, hu2014reliability} have been proposed with the aim of improving the resilience of the control plane with respect to unexpected network failures. For example, the research work \cite{hu2017reliable} developed a reliable controller placement strategy taking both the node reliability and link quality into consideration.

Nevertheless, a majority of existing works on the CPP aim to improve network performance without explicitly considering the impact of varied workload distributions among controllers. Particularly, the workload on each controller is usually oversimplified to be identical \cite{hock2013pareto}. In some extreme cases \cite{controllerPlacement}, the controller workload is completely ignored. 
However, in reality, the workload distribution can bring significant changes to the controller performance. Specifically, when the controller workload reaches a certain level, the processing time can increase substantially \cite{yao2014capacitated}. 
In this case, the overall response time of a controller mainly depends on the processing time while the impact of communication delay becomes trivial.



In this paper, we will investigate the CPP based on our recently proposed BindingLess Architecture for Distributed Controllers (BLAC) \cite{huang2017blac}. BLAC introduces a scheduling plane to 
allow dynamic association between any controllers and any switches. In other words, every switch enjoys the flexibility of passing its processing requests to arbitrary controllers available in the network. This is vital for us to explicitly control the workload distribution across all controllers.

Based on BLAC, the CPP can be addressed in a systematic manner. Specifically, a queuing model will be established in this paper to simultaneously consider the impact of controller workload distribution, communication delay and control plane utilization on network performance. Unlike existing works restricted by the assumption of equal workload over all controllers, we consider a more realistic scenario in which the workload distribution matches closely with the controller location, capacity and other network settings. 
For example, in a heterogeneous network setting (i.e., controllers with different processing capacities are located in various geographic locations, leading to significantly different network latency), dispatching workload evenly is unlikely to achieve reasonable performance. Alternatively, sending more requests to nearby controllers without overloading them can be a better option. 
Our new model of the CPP quantifies the benefits of the latter option and describes the CPP in the form of a constrained optimization problem with the objective of  
simultaneously reducing the packet response time and improving the control plane utilization.  

Given the fact that CPPs without taking controller workload distribution into account are already NP-hard \cite{controllerPlacement,bo2016controller,yao2014capacitated}. The CPP considered in this paper is also NP-hard since it generalizes previous CPPs by supporting dynamically changing controller workload.
To tackle this problem, a new direct optimization method is designed by combining the use of both the genetic algorithm (GA) and gradient descent (GD) optimization to obtain near-optimal solutions. Particularly, GA is used to optimize both the number and locations of controllers in the network, according to which GD is further exploited to optimize the corresponding workload distribution over all deployed controllers. 


In previous works \cite{arab2015adaptive,albero2006combined,li2013hybrid}, GA and GD are combined to build a memetic algorithm. In such an algorithm, evolutionary search is used to explore new solutions while GD-based local search is used for improving existing solutions. For example, \textit{Li et al.} \cite{li2013hybrid} solved a microphone array placement problem by applying GA to find a candidate solution which was further improved by GD. Different from these works, GD in our algorithm is used purely for fitness evaluation (no existing solutions will be improved through GD). Although we realize that similar approaches have been spotted lately in designing artificial neural networks \cite{liu2018structure,real2017large}, our algorithm, however, aims to solve a very different problem in SDN. To the best of our knowledge, we are the first to simultaneously use GA and GD to address such networking problems.

To evaluate the performance of our proposed method, extensive simulations have been conducted under two network settings in terms of controller specification and network sizes. Simulation results show that our approach can effectively strike the trade-off between the controller utilization and response time.


\section{Related Work} \label{sec:relatedwork}
Existing studies on the CPP can be generally classified into two categories: uncapacitated CPP (UCPP) and capacitated CPP (CCPP). Literature \cite{controllerPlacement, hu2013reliability, hu2014reliability,obadia2015greedy} in UCPP barely considers the workload distribution among controllers when solving the CPP. For example, when the CPP is first proposed \cite{controllerPlacement}, the problem is assumed to be static and formulated as a K-median problem with the aim of minimizing the average communication delay among switches and controllers, which is further showed as NP-hard. \textit{Obadia et al.} \cite{obadia2015greedy} formulated the CPP as the control plane overhead minimization problem and solved by a greedy heuristic. 
Note that these works all assume that the controller processing time is negligible compared with the communication delay. However, this assumption is not always valid. For example, when the controller workload reaches a certain level, the processing time can increase substantially which turns out to be a dominating factor in the packet response time \cite{yao2014capacitated}.

Recently, an increasing number of research works \cite{hock2013pareto, lange2015heuristic,yao2014capacitated} realize the importance of considering the workload distribution among deployed controllers.
\textit{Yao et al.} \cite{yao2014capacitated} redefined the CPP as a capacitated controller placement problem which considered the controller capacity as a constraint. 
With the aim of minimizing the load diversity among controllers, \textit{Hu et al.} \cite{bo2016controller} applied GA to divide the network into a given number of domains so as to balance the workload in each domain.
Similar research can also been found in \cite{ksentini2016using,ul2015revisiting,hock2013pareto}. Despite their promising performance, these methods tend to oversimplify the workload distribution. For instance, \textit{Hock et al.} \cite{hock2013pareto} considered that the workload of a controller equaled the number of connected switches. In other words, all switches are assumed to receive the same amount of network traffic. \textit{Huque et al.} \cite{ul2015revisiting} set the controller workload as a fixed value corresponding to the maximum requests it can receive from the switches. 
However, these assumptions are unlikely to remain valid since the network traffic is highly dynamic and unpredictable. Therefore, how to model the impact of workload distribution in the CPP remains unsolved. 

Apart from that, existing works on the CPP are based on static-binding-based controller architectures which can easily lead to controller overloading when associated switches generate high volume of requests within a short time frame \cite{elasticon}. Although dynamic-binding-based controller architectures \cite{elasticon} have been proposed to alleviate this issue by migrating switches from overloaded controllers to underloaded ones, the requests from one switch can still only be processed by one controller. Thus, the granularity of workload distribution is still limited to switch level. In this case, the destination controller is prone to be overloaded since its workload will increase dramatically if the switch migrated to it has accumulated a large number of pending requests. 

To enable fine-grained workload distribution, we will investigate the CPP based on our recently proposed architecture BLAC. BLAC introduces bindingless switch-controller association so that requests from one switch can be flexibly processed by different controllers. The detailed design of BLAC will be demonstrated in Section \ref{sec:blac}. On top of BLAC, the CPP is formulated as an optimization problem aiming at simultaneously reducing the response time while maintaining high control plane utilization. Different from previous works that tend to ignore or oversimplify the workload distribution, the relationship among the network response time, the workload distribution and communication delay will be systematically captured using a queuing model. Apart from that, we investigate the combined use of GA and GD to simultaneously identify both the CPP solution and the optimal workload distribution. 

\section{BindingLess Architecture for Distributed Controllers} \label{sec:blac}
In this section, we briefly review BLAC and its design. After that, key reasons for choosing BLAC in this study will be elaborated. 

\subsection{Architecture Design}
\begin{figure}[!tbp]
\centering
\includegraphics[width=0.9\linewidth]{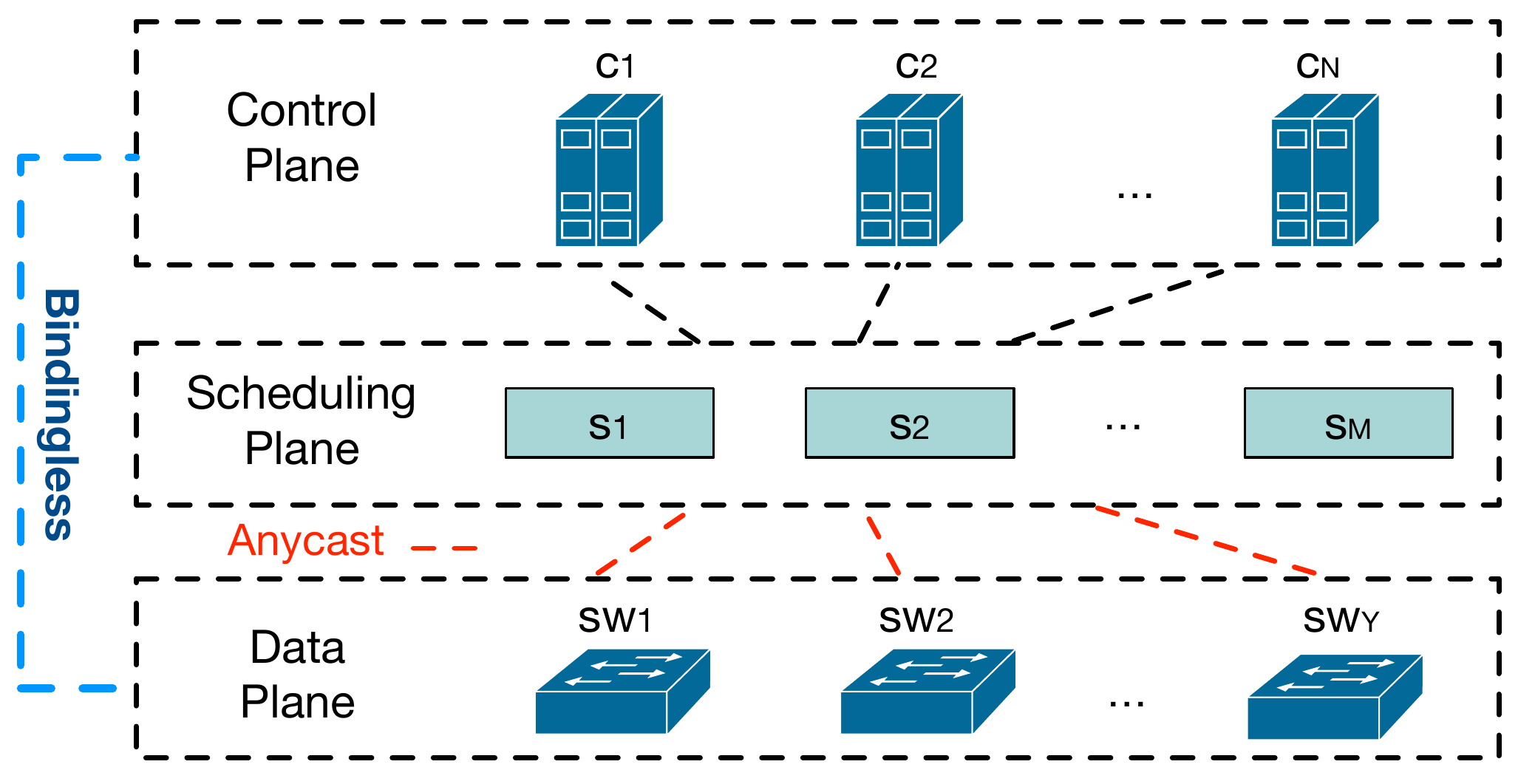}
\caption{System Architecture.}
\label{fig:architecture}
\vspace{-10pt}
\end{figure}

As shown in Fig. \ref{fig:architecture}, BLAC is comprised of three components: the data plane made up of numerous switches and communication links, the scheduling plane consisting of multiple schedulers, and the control plane composed of distributed controllers that jointly provide consistent network control.

The switches, denoted as $\boldsymbol{sw}=\{sw_1, sw_2, ..., sw_{Y}\}$ in the data plane, are in charge of packet forwarding based on well-defined flow rules. Specifically, upon receiving a packet, $sw_y$ will try to match the packet against any rules stored in its Ternary Content Addressable Memory (TCAM) \cite{tcam}. Only if no matching rule can be identified, a routing request will be forwarded to the scheduling plane for subsequent processing. 

The scheduling plane is supported by a group of $M$ schedulers, i.e., $\boldsymbol{s} = \{s_1, s_2, ..., s_M\}$. Each scheduler communicates with switches through the anycast technique \cite{anycast}, which guarantees low network latency and simplifies network configuration. Specifically, the switch will automatically connect to the nearest scheduler, reducing the network communication delay. Besides, all schedulers listen on a unique anycast address for the requests from nearby switches, enabling simple switch configuration. Note that the schedulers are low-cost and easy-to-deploy NFV entities. They can be strategically placed in many locations of a network to minimize communication delay without incurring too much cost. Thus, the scheduling plane should never become the bottleneck of the architecture. It is therefore not important for us to treat scheduler placement as a separate problem.

On top of the scheduling plane in Fig. \ref{fig:architecture} is the control plane, where $N$ distributed controllers, denoted as $\boldsymbol{c} = \{c_1, c_2, ..., c_N\}$, will be established according to the controller placement decisions. Specifically, the processing capacities of the $N$ controllers can be captured through vector $\boldsymbol{\alpha} = (\alpha_1, \alpha_2, ..., \alpha_N)\in\mathrm{R}^N$ where $\alpha_n$ represents the maximum number of routing requests that controller $c_n$ can process within a unit of time (i.e., a second). 
Each controller is free to communicate with any scheduler in the scheduling plane. The communication delay between $\boldsymbol{s}$ and $\boldsymbol{c}$ is collectively expressed through a matrix $\boldsymbol{D}\in \mathrm{R}^{M \times N}$ where $d_{mn}$ is the communication delay between $s_m$ and $c_n$. Every time when a scheduler $s_m$ receives a packet (or routing request), it must forward it to one of the controllers for processing. Such a controller $c_n$ will be selected according to a probability denoted as $p_{mn}$. For concise discussion, we jointly present the selection probabilities for all controllers in the form of a matrix $\boldsymbol{P} \in \mathrm{R}^{M \times N}$.  

\subsection{Using BLAC for the Controller Placement Problem}
In this research, we decide to use BLAC to address the CPP because of two main reasons:

First, BLAC enables dynamic and transparent controller placement without incurring time-consuming switch migration. Other distributed controller architectures such as ONOS \cite{onos} and ElastiCon \cite{elasticon} support either static or dynamic switch-controller binding. Therefore, when the controllers are relocated, switches need to migrate from previous controllers to new ones. This process can be time-consuming and even introduce network disruption. On the contrary, BLAC introduces bindingless switch-controller association. In BLAC, switches are connected directly to the scheduling plane. Thus, controllers can be flexibly and transparently relocated without interrupting the switch connections. 

Second, BLAC enables fine-grained controller load control. Due to the switch-controller binding constraint, the majority of distributed controller architectures can only manipulate the controller workload at switch level. In other words, requests from one switch can only be processed by one controller. However, BLAC can dispatch requests from one switch to different controllers, achieving flow-level load control. Thus, the architecture improves the flexibility of distributing the workload among controllers.

Apart from providing flexible and explicit control over the controller workload distribution, the complete separation between the control plane and the data plane also enables us to formulate the CPP without the switch-controller binding constraint. With the help of BLAC, the CPP is effectively formulated in Section \ref{subsection:controller_placement_problem_formulation}.



\section{Controller Placement} \label{sec:placement}
After reviewing the architecture design, we now consider how to solve the CPP based on BLAC.
In this section, we address the CPP for the purpose of improving network response time and control plane utilization. 
Particularly, we can formulate the CPP as a constrained optimization problem using a queuing model.
In line with the problem formulation, we will further investigate several alternative ways to solve the CPP, ranging from simple heuristic approaches to more advanced search methods.

\subsection{Problem Formulation} \label{subsection:controller_placement_problem_formulation}

Using BLAC introduced in Section \ref{sec:blac}, within a reasonable period of time, we assume that in the queuing model, the requests arriving at $s_m$ ($m=1,...,M$) follow a Poisson distribution with a stable arrival rate denoted as $ \lambda_m$.
This is a common assumption in the literature  \cite{bai2017pias,guo2017pricing,wang2016dynamic}, enabling us to easily analyze the performance of controllers. 
Although our formulation is based on this assumption, we measured the simulation results under different traffic distribution and found that the simulated performance did not depend heavily on this assumption.
As previous work \cite{huang2017blac,wang2016dynamic}, we also assume that the complexity for processing each request by the same controller is roughly identical since the complexity of routing decisions depends on the network size that remains stable within a long time span \cite{dijkstra1959note}. While the complexity of routing decisions being identical, it should also be noted that if controllers have different capacities, the processing time would vary significantly from one controller to another. 

Given a group of candidate controllers $\boldsymbol{c}$ with fixed locations selected based on network constraints (e.g. bandwidth),
the job for the CPP is to select a subset of locations (controllers) to meet the performance demand from the data plane. In view of this, the CPP can be converted to a controller location selection problem. Note that different from previous works \cite{controllerPlacement,hu2013reliability,yao2014capacitated} which assume the number of selected controllers is given, the CPP to be tackled here requires the number of deployed controllers to be determined too. In particular, the solution of the CPP can be represented as a binary vector $\boldsymbol{x} = (x_1,x_2,...,x_N) \in \mathrm{R}^N$, where each dimension of $\boldsymbol{x}$ can be represented as
\begin{equation}
    x_n = \begin{cases}
 \; 1, \text{ if } c_n \text{ is selected} \\ 
 \; 0, \text{ otherwise }
\end{cases}
\end{equation}
and the number of selected controllers can be calculated as ${\sum}_{n=1}^N x_n$. 
The key notations are summarized in Table \ref{notation} for the ease of reference.

\begin{table}[]
\centering
\caption{Mathematical notations}
\label{notation}
\begin{tabular}{c|l}
\hline
Notation                      & Definition          \\ \hline
$sw_y$                        & The $y^{th}$ switch\\
$s_m$                         & The $m^{th}$ scheduler \\
$c_n$                         & The $n^{th}$ controller\\
$M$                           & Number of schedulers                    \\ 
$N$                           & Number of controllers                   \\
$\lambda_m$                   & Request arrival rate of the $m^{th}$ scheduler\\
$\alpha_n$                    & Processing capacity of the $n^{th}$ controller\\
$\beta_n$                     & Decay factor of the $n^{th}$ controller\\
$x_n$                         & Whether the $n^{th}$ controller is selected\\
$d_{mn}$                      & Communication delay between the $m^{th}$ scheduler and \\
& the $n^{th}$ controller\\
$p_{mn}$                      & Probability of the $m^{th}$ scheduler sending requests to \\
& the $n^{th}$ controller \\\hline
\end{tabular}
\vspace{-4mm}
\end{table}


Similar to recent works \cite{mahmood2015modelling,wang2016dynamic} which try to understand the controller's behavior using a queuing model, we consider each controller as an independent M/M/$1$ queue. Unlike these works which model the controller traffic with the switch-controller binding constraint, we consider that each controller maintains a queue for pending requests arriving from any scheduler/switch.
This assumption facilitates simple and efficient implementation of schedulers (no queuing management required). Accordingly, the workload of $c_n$ can be determined as:
\begin{equation}
\label{eqt:placement_controller_workload}
\theta_n = \sum_{m=1}^{M}\lambda_mp_{mn}
\end{equation}

According to Little's Law \cite{little'slaw}, the average processing time of $c_n$ with capacity $\alpha_n$ is:
\begin{equation}
\label{eqt:placement_controller_processing_time}
\upsilon _n=\frac{1}{\alpha_n-\theta_n}
\end{equation}

Since the network we consider here may span large geographic areas, the communication delay $d_{mn}$ between $s_m$ and $c_n$ must be taken into account when computing the response time. The average communication delay of $c_n$ can be represented by: 
\begin{equation}
\label{eqt:placement_network_latency}
l _n=\frac{\sum_{m=1}^{M}\lambda_mp_{mn}d_{mn}}{\theta_n}
\end{equation}

In consideration of both the processing time and the communication delay, the average response time of $c_n$ can be calculated as: 
\begin{equation}
\label{eqt:placement_response_time}
t_n=\upsilon_n + 2l_n
\end{equation}

Based on the design of BLAC, it is easy to see that the communication delay between a switch and the scheduler responsible for handling its requests is generally fixed. We can also see that the time required for a scheduler to dispatch a request to any controller is negligible (provided that a time-efficient scheduling algorithm is used). Therefore in our CPP, we only need to consider $t_n$ that measures the response time between schedulers and controllers.

To simplify our discussion, our model takes no consideration of synchronization costs among controllers. The simplification is justifiable due to the following aspects: (1) State-of-the-art distributed controller architectures/frameworks are designed to incorporate a large number of controllers such that multiple optimization techniques (e.g., the anti-entropy mechanism \cite{onos}) have been applied to achieve almost negligible synchronization costs. (2) Even if the costs cannot be ignored, their impact can be easily incorporated into our model by adjusting the controller capacities accordingly. Based on the above arguments, a majority of existing research works \cite{controllerPlacement,hu2013reliability,yao2014capacitated} tend to safely ignore the synchronization costs.

The average response time over all requests generated by the data plane can be decided from below:
\begin{equation}
\label{eqt:placement_avg_response_time}
t= \frac{\sum_{n=1}^{N}\theta_nt_n}{\sum_{n=1}^{N}{\theta_n}}
\end{equation}

Given the request arrival rate and the controller selection decision $\boldsymbol{x}$, the controller utilization can be determined as:
\begin{equation}
\label{eqt:placement_avgcontroller_utilization}
u=\frac{\sum_m^M{\lambda_m}}{\sum_n^N \alpha_nx_n }
\end{equation}

In this study, we choose to use $t$ in (\ref{eqt:placement_avg_response_time}) as the metric for control plane performance. In fact, by selecting controller locations properly, the performance of the control plane can be improved, resulting directly in reduced $t$. Meanwhile, to maintain reasonable operational cost, the average controller utilization $u$ must be kept at a high level to avoid controller over-provisioning. In consideration of both requirements, an intuitive approach is to formulate the CPP as below:
\begin{equation} \label{eqt:intuitive_cpp_formulation}
    \underset{\boldsymbol{x}\in \{0,1\}^N}{\text{minimize}} \quad
 \underset{\boldsymbol{P}}{\text{minimize}} \quad \mu_t \; t(\boldsymbol {x},\boldsymbol{P})-\mu_u \; {u(\boldsymbol {x})}
\end{equation}
Note that both $t(\boldsymbol {x},\boldsymbol{P})$ and $u(\boldsymbol{x})$ are direct functions of placement vector $\boldsymbol{x}$ and workload distribution probability $\boldsymbol{P}$. $\mu_t$ and $\mu_u$ represent the weight coefficients which strike the trade-off between the response time and utilization. However, this formulation may not be suitable for two reasons. First, it requires a high level of domain expertise to select the appropriate value for the weights. Second, \eqref{eqt:intuitive_cpp_formulation} requires both objective values to be normalized.

In consideration of the above reasons, the CPP is formalized as below: 
\begin{align}
\label{eqt:placement_obj}
& \underset{\boldsymbol{x}\in \{0,1\}^N}{\text{minimize}}
& & \underset{\boldsymbol{P}}{\text{minimize}} \quad \frac{t(\boldsymbol {x},\boldsymbol{P})}{u(\boldsymbol {x})} \\
& \text{subject to}
\label{eqt:placement_constraint1_capacity}
& & \theta_n\leq \beta_n \alpha_n, \; \forall n \\
\label{eqt:placement_constraint2_probability}
&&& 0\leq p_{mn}\leq 1, \; \forall m, n \\
\label{eqt:placement_constraint3_probabilitysum}
&&& \sum_{n=1}^N p_{mn}=1, \; \forall m 
\end{align}
Inequality (\ref{eqt:placement_constraint1_capacity}) guarantees that the workload of each controller chosen by $\boldsymbol{x}$ will not exceed their capacity. Specifically, through properly adjusting $\boldsymbol{\beta}=(\beta_1, \beta_2, ..., \beta_N)$, sufficient capacity can be reserved at each controller to cope with unexpected ephemeral traffic bursts. Both inequality (\ref{eqt:placement_constraint2_probability}) and equality (\ref{eqt:placement_constraint3_probabilitysum}) jointly ensure that the packet distribution probabilities $p_{mn}$ are well-defined.

\subsection{Placement Optimization} \label{subsection:placement optimization}
The CPP can be considered as a variant of the facility location problem, which is known as NP-hard \cite{controllerPlacement}. Thus, finding the optimal solution is almost impossible for large networks. To solve this problem efficiently, we will study and compare several alternative approaches.

\subsubsection{\textbf{Random Approach}}
One simple and widely used heuristic is the random selection. It selects controllers randomly and uniformly from the given set of candidate controllers $\boldsymbol{c}$ until total capacity of all selected controllers reaches a given level (e.g., $\gamma\left \| \boldsymbol{\lambda} \right \|_1 $, where $\gamma$ is a provisioning factor and $\left \| \boldsymbol{\lambda} \right \|_1$ is the $L_1$ norm of request arrival rate $\boldsymbol{\lambda}$). Due to its simplicity, the random approach is widely utilized in real-world networks \cite{joo2009performance,liang2008random}. However, the effectiveness of the random approach relies on the assumption that all controllers have identical capacities and propagation latencies. It is unlikely for this assumption to remain valid with SDN controllers hosted by hybrid data centers.

\subsubsection{\textbf{Capacity-based Greedy Approach}}\label{subsubsec:greedy}
To address the limitations of the random approach, a greedy approach is considered. Specifically, the controller selection is performed in an iterative manner. In each iteration, the controller with the highest capacity among existing unselected controllers is chosen. The iteration terminates when the capacity of all selected controllers reaches a certain level. 


Although the greedy approach is simple and efficient, it only considers the controller capacity, deliberatively ignoring the utilization of the control plane and the communication delays between controllers and schedulers. As a result, it may lead to high operational costs and long communication delays.


\subsubsection{\textbf{K-median Approach}} \label{subsubsec:k_median}
Currently, one of the most widely-adopted controller placement strategies is called K-median introduced by \textit{Heller et al.} \cite{controllerPlacement}. K-median aims to find the controller locations that minimize the average communication delay. Note that the number of controllers is generally assumed to be given or can be easily obtained in the conventional K-median approach. However, finding the number of controllers that can closely match the network traffic in a wide-area network equipped with controllers with various capacities can be complicated. Thus, in this paper, we drop the assumption and adapt K-median to solve the CPP by adding one constraint, i.e., the total capacity of all selected controllers must reach a given level. The adapted K-median can be formulated as:
\begin{align}
\label{eqt:k-median}
& \underset{x_n\in \{0,1\}}{\text{minimize}}
& & \frac{\underset{m,n}{\sum} d_{mn} x_n}{\underset{n}{\sum} x_n} \\
& \text{subject to}
& & \underset{n}{\sum} x_n \alpha_n \geq \gamma\left \| \boldsymbol{\lambda} \right \|_1
\end{align}
where $\gamma$ and $\left \| \boldsymbol{\lambda} \right \|_1$ are the provisioning factor and total request arrival rate described in random approach.

To solve this problem, an $n$-approximation algorithm called forward greedy approach is adapted\cite{kMedian_forwardGreedy}. It starts by assigning a location $x_n$ with smallest $d_{m,n}$ to be $1$. Then at each iteration, the controller with the least average communication delay among existing unselected controllers is chosen until the total capacity constraint is satisfied.

Compared with previous two heuristics, K-median approach focuses on minimizing the communication delay which plays an important role in network response time. Therefore, the K-median approach is expected to be more effective than previous heuristics, especially in wide-area networks. However, the K-median approach still cannot effectively manage the utilization of the control plane, which may result in additional operating costs.

\setlength{\textfloatsep}{10pt}
\begin{algorithm}
  \caption{GD-based Fitness Evaluation} \label{alg:evaluation}
  \begin{algorithmic}[1]
    \STATE Initialize the workload distribution $\boldsymbol{P}$ 
    \FOR {iteration $= 1, 2, ... $}
      \STATE Calculate the gradient $\boldsymbol{g}$ using Theano: $\boldsymbol{g} \leftarrow \frac{\partial f}{\partial \boldsymbol{P}}$  
      \STATE Update $\boldsymbol{P}$ based on \eqref{eqt:parameter_update}
    \ENDFOR
    \RETURN Fitness value of a given solution $\boldsymbol{x} \leftarrow f(\boldsymbol{x},\boldsymbol{P})$
  \end{algorithmic}
\end{algorithm}

\subsubsection{\textbf{Direct Optimization Approach}} \label{subsubsec:ga}
In literature, evolutionary computation (EC) algorithms are often exploited to find near-optimal solutions to NP-hard problems \cite{0/1knapsack_problem,nsga,eiben2015evolutionary}. They stand for a promising alternative approach to tackle the CPP formulated in \eqref{eqt:placement_obj}. Specifically, GA with a binary solution representation is employed by us for controller placement. GA is an EC approach inspired by Darwin's theory of evolution. It solves the optimization problem by simulating the process of natural selection to find highly fit solutions to a given problem. In this paper, we follow the GA framework introduced in \cite{GA_tutorial}. Each individual in a GA population is a binary vector $\boldsymbol{x}$ that directly represents a solution to our CPP, as explained in Section \ref{subsection:controller_placement_problem_formulation}. 

Given any possible controller placement solution $\boldsymbol{x}$, it must satisfy all constraints from \eqref{eqt:placement_constraint1_capacity} to \eqref{eqt:placement_constraint3_probabilitysum}. To enforce constraint satisfaction during the evolution, we transform the constrained optimization problem in Section \ref{subsection:controller_placement_problem_formulation} into an unconstrained one by introducing several penalty terms \cite{constraint_handling_stochastic}. Consequently, the fitness function for GA to determine the goodness of solution $\boldsymbol{x}$ becomes:
\begin{equation}
\begin{split}
\label{eqt:placement_fitness}
f(\boldsymbol{x}) & = \underset{\boldsymbol{P}}{\text{minimize}} \{ \frac{t(\boldsymbol{x},\boldsymbol{P})}{u(\boldsymbol{x})}\\
& \quad -\mu_1\sum_{n=1}^{N} \text{min}(0,\beta_na_n-\theta_n) \\
& \quad -\mu_2\sum_{m=1}^M \text{min}(0,1-\sum_{n=1}^{N-1} p_{mn}) \\
& \quad -\mu_3\sum_{m=1}^M\sum_{n=1}^N \text{min}(0, p_{m,n}) \}
\end{split}
\end{equation}
where $\mu_1$, $\mu_2$, and $\mu_3$ are penalty coefficients used to quantify the imposed penalty terms. 
Usually, the value of penalty coefficients can be set as a monotonically decreasing function of the generation number in GA so that infeasible solutions are greatly penalized and the existing solutions in a GA population are forced to satisfy the constraints. 

According to \eqref{eqt:placement_fitness}, the fitness of $\boldsymbol{x}$ is obtained through minimizing \eqref{eqt:placement_fitness} with respect to $\boldsymbol{P}$. Such minimization can be approached straightforwardly through a gradient descent (GD) approach, as shown in Algorithm \ref{alg:evaluation}. In each iteration of GD, given candidate solution $\boldsymbol{x}$ for our CPP, the gradient $\boldsymbol{g}$ of the fitness function $f(\boldsymbol{x}, \boldsymbol{P})$ with respect to $\boldsymbol{P}$ is calculated via Theano \cite{theano}, an efficient gradient computation tool for large-scale optimization. Instead of updating the parameter $\boldsymbol{P}$ using a constant learning rate $l$, we adaptively adjusts $l$ from $l_H$ (a higher threshold) to $l_L$ (a lower threshold) using equation \eqref{eqt:learning_rate_update}:
\begin{equation}\label{eqt:learning_rate_update}
l = l_H - \frac{(l_H-l_L)}{N_I}i
\end{equation}
where $N_I$ is the total number of iterations and $i$ represents the $i^{th}$ iteration.

Based on $l$ in \eqref{eqt:learning_rate_update} and the gradient $\boldsymbol{g}$ calculated by Theano, $\boldsymbol{P}$ can be updated as:
\begin{equation}\label{eqt:parameter_update}
\boldsymbol{P}_{i+1} =\boldsymbol{P}_i - l\boldsymbol{g} 
\end{equation}
The whole process iterates until the stopping criteria are reached. As a result of minimizing \eqref{eqt:placement_fitness} through iterative updating $\boldsymbol{P}$ in GD, the fitness of $\boldsymbol{x}$ can be obtained eventually. 



\setlength{\textfloatsep}{10pt}
\begin{algorithm}
      \caption{Direct Optimization Algorithm (GA+GD)}
      \label{alg:gagd}
       \begin{algorithmic}[1]
          \STATE Initialize the population
          \STATE Evaluate the population using Algorithm \ref{alg:evaluation}
          \FOR {generation $ = 1, 2, ... $}
          	\STATE Generate new individuals via Crossover and Mutation
          	\STATE Evaluate the new individuals using Algorithm \ref{alg:evaluation}
          	\STATE Select individuals to form a new population 
          	
          
          \ENDFOR
          \RETURN The best individual
       \end{algorithmic}
\end{algorithm}

Based on the fitness values obtained from Algorithm \ref{alg:evaluation}, the best performing individuals are selected as the basis for the new generation that is produced via genetic operators (e.g., mutation and crossover). This evolution process repeats over many iterations until a sufficiently fit solution is found. The overall procedure is summarized in Algorithm \ref{alg:gagd}. 

\section{Evaluation} \label{sec:evaluation}
In this section, we first present the evaluation setup. Then the performance of our algorithm will be examined using two representative networks. To ease the discussion, we start with a small-scale network. To generalize our simulation setting, we consider that the network is equipped with a set of heterogeneous SDN controllers. Such a setting is frequently seen especially for small data centers hosted on the premises of many organizations in the process of hardware upgrading. After that, we consider a large-scale network that supports a hybrid data center with controller resources scattered around the globe. In this case, the CPP becomes extremely challenging (i.e., NP-hard) making simple heuristics ineffective. On the other hand, GA is expected to address this complex problem effectively.


\subsection{Evaluation Setup} \label{subsec:exp_setup}
During the evaluation, the network topology we used is the fat-tree topology
which is widely used in real-world data centers (e.g., Facebook \cite{facebook_fat_tree} and Google \cite{fat-tree}) due to its simplicity and efficiency. 

Note that the communication delay highly depends on the geographic distance covered by the network. We simulate the communication delay differently in two networks. In particular, for the small-scale network, we consider that all network devices are placed within the data center. 
Therefore the communication delay can be safely ignored (below $1$ms) \cite{guo2015pingmesh}. Nonetheless, to make the simulation more accurate, we simulate the latency sampled from a distribution observed in a real-world data center \cite{guo2015pingmesh}. For the large-scale network which may span large geographic areas, the communication delay becomes significant. According to existing studies, the communication delay can vary from $10$ ms to $200$ ms \cite{he2013next,choy2012brewing,li2010cloudcmp}. In our simulations, the communication delay is simulated by sampling from a distribution measured in \cite{li2010cloudcmp}.   


Regarding the parameter tuning, we mainly follow the typical settings in existing works \cite{srinivas1994genetic,srinivas1994adaptive}. Specifically, a chromosome in GA is a binary vector representing a CPP solution. The mutation and crossover rates are $0.1$ and $1$ respectively. For fitness evaluation, GD is performed for $15$ iterations so as to balance the algorithm efficiency and performance effectiveness. As for the penalty coefficients $\boldsymbol \mu=(\mu_1,\mu_2,\mu_3)$ in \eqref{eqt:placement_fitness}, we set them to be $10$ for all simulations. In fact, different settings of $\boldsymbol \mu$ ranging from $1$ to $100$ have been tested and no significant performance impact was noticed. We therefore believe our algorithm can work robustly with varied settings of $\boldsymbol\mu$. Besides, the decay factor $\boldsymbol\beta$ in \eqref{eqt:placement_fitness} for each controller is set to be $0.83$. Correspondingly, the provisioning factor $\gamma$ for the heuristics is set to be $1.2$. 

\begin{table*}[!tbp]
\centering
\caption{Controller placement performance comparison in a small-scale network with heterogeneous controllers.}
\label{tab:scen2_placement_comparison}
\resizebox{1\textwidth}{!}{
\setlength{\tabcolsep}{3pt}
\begin{tabular}{|c|c|c|c|c|}
\hline
\textbf{CPP Algorithm} & \textbf{Random Approach} & \textbf{Capacity\_based Greedy Approach} & \textbf{K-median Approach} & \textbf{GA+GD} \\ \hline
\textbf{Average Response Time (ms)} & 0.3928 & \textbf{0.2807} & 0.2816 & 0.2816 \\ \hline
\textbf{Control Plane Utilization (\%)} & \textbf{80.00} & 66.67 & 72.72 & \textbf{80.00} \\ \hline
\textbf{Controller Number} & \textbf{4} & \textbf{4} & 5 & \textbf{4} \\ \hline
\end{tabular}
}
\vspace{-2mm}
\end{table*}

\subsection{Small-scale Network} \label{subsec:small-dc-diff-ctl}

The topology consists of $80$ switches at a scale comparable to many university and enterprise data centers \cite{networktraffic}. We set up a group of $10$ candidate controllers with different capacities. Since the network we used in this setting is small, we only consider controllers with two different capacities (i.e., $45,000$ and $30,000$ pkt/s). The total number of requests generated by the entire data plane is averaged to be $120,000$ pkt/s. 

We measured and compared the network performance with different numbers of schedulers. The results show that there is no significant difference in terms of the network performance. This finding confirms our belief that schedulers are not the performance bottleneck. Given the relatively small network size, we decide to report results when only one scheduler is used. In the small-scale network setting, we first demonstrate the influence of workload distribution on network performance. After that, we evaluate the performance of different placement methods mentioned in Section \ref{subsection:placement optimization} using BLAC.


To demonstrate that it is important to consider workload distribution, comparison experiments are conducted. Specifically, given a placement solution, we distribute requests using the workload distribution optimized by GD and measure the network response time. In comparison, without the help of GD, we consider that each switch is statically connected to its nearest controller and sends all requests to that controller. In terms of CPP strategies, we run experiments using different heuristics (e.g., greedy and K-median) and all of them show similar performance patterns. Due to the space limit, we only report results with K-median.

\begin{figure}[!tbp]
\centering
\includegraphics[width=0.7\linewidth]{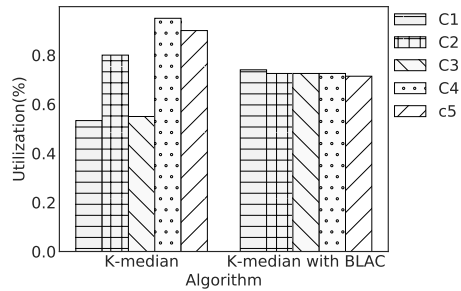}
\caption{Controller utilization comparison between K-median and K-median with GD.}
\label{fig:utilization}
\vspace{-10pt}
\end{figure}

As we expected, the average response time we measured without considering workload distribution is $0.65$ ms in K-median which is over $2$ times larger than $0.28$ ms in K-median+GD. This is because K-median solely optimizes the communication delay without considering the controller workload distribution. 
Note that in a small-scale network with negligible communication delay, it is preferable to dispatch requests to controllers so that the utilization is balanced among all controllers.
Since the switches and network traffic are not evenly distributed within the network, without considering workload distribution can easily lead to uneven control plane utilization. 
To verify this, we measure the controller utilization as the total requests received by the controller divided by the controller capacity. As shown in Fig. \ref{fig:utilization}, with the help of GD, the controller utilization is relatively even and less than $80\%$. On the other hand, we also notice that in the case that only K-median is used, $2$ controllers are highly-loaded (i.e., over $90\%$ utilization)  while $2$ controllers remain under-utilized (i.e., less than $60\%$ utilization), rendering comparatively high response time.          

To compare the performance of different placement methods mentioned in Section \ref{subsection:placement optimization}, two performance metrics including the average response time defined in \eqref{eqt:placement_avg_response_time} and control plane utilization defined in \eqref{eqt:placement_avgcontroller_utilization} are evaluated. In particular, regardless which algorithm is used to deploy controllers in the following simulations, the workload distribution over all deployed controllers will be consistently optimized through GD.

Intuitively, given the same request arrival rate, controllers with more capacities tend to achieve less response time. Consequently, we expect the capacity-based greedy heuristic to be highly effective. Our simulation result as shown in Table \ref{tab:scen2_placement_comparison} clearly demonstrates the usefulness of this greedy heuristic with $0.2807$ ms response time. However, we notice that GA provides an alternative selection which provides similar response time ($0.2816$ ms) with higher control plane utilization ($80\%$).
Although choosing the most powerful controllers as capacity-based greedy approach can clearly achieve best response time, it has resulted in low control plane utilization (e.g., $66.7\%$ in this example), incurring additional energy and operational costs. With regard to GA, it achieves clearly better trade-off between response time and controller utilization, thus is more suitable for controller placement. 

\begin{table*}[!tbp]
\centering
\caption{Controller placement performance comparison in a large-scale network with heterogeneous controllers.}
\label{tab:scen3_placement_comparison}
\resizebox{1.0\textwidth}{!}{
\setlength{\tabcolsep}{3pt}
\begin{tabular}{|c|c|c|c|c|}
\hline
 \textbf{CPP Algorithm} & \textbf{Random Approach} & \textbf{Capacity\_based Greedy Approach} & \textbf{K-median Approach} & \textbf{GA+GD} \\ \hline
\textbf{Average Response Time (ms)} & 16.3 & 26.1 & \textbf{8.5} & \textbf{8.5} \\ \hline
\textbf{Control Plane Utilization (\%)} & 81.36 & 80.00 & 76.19 & \textbf{82.75} \\ \hline
\textbf{Controller Number} & 18 & \textbf{10}  & 20 & 16 \\ \hline
\end{tabular}
}
\vspace{-4mm}
\end{table*}

\subsection{Large-scale Network} \label{subsec:large-network-diff-ctl}

To evaluate the performance of GA, we simulate a network with a total number of $720$ switches, at a scale comparable to many large commercial data centers \cite{networktraffic}. 
Besides, $50$ controller candidates are provided for the CPP with capacities spanning from $15,000$ pkt/s to $90,000$ pkt/s \cite{kreutz2015software}.

Table \ref{tab:scen3_placement_comparison} shows the evaluation result using different placement methods with the combined request arrival rate over all switches in the network reaching $720,000$ pkt/s. Significantly different from Section \ref{subsec:small-dc-diff-ctl}, capacity-based greedy approach has the highest response time ($26.1$ ms) among all methods. This is understandable because the communication delay plays a major part in the response time in a large-scale network while capacity-based greedy heuristic chooses controllers solely based on their capacities. Thus, remote controllers may be selected, resulting in large response time. We also notice that the K-median approach is a good competitor to GA with respect to response time ($8.5$ ms for both GA and K-median). 
However, in terms of controller utilization, GA with $82.75\%$ utilization outperforms the K-median approach with $76.19\%$ utilization thanks to our design of fitness function in (\ref{eqt:placement_fitness}) that enables GA to optimize both the response time and controller utilization simultaneously.

\begin{figure}[!tbp]
  \begin{center}
    \subfloat[Response time]{\label{scen3_resp_time}\includegraphics[width=.5\linewidth]{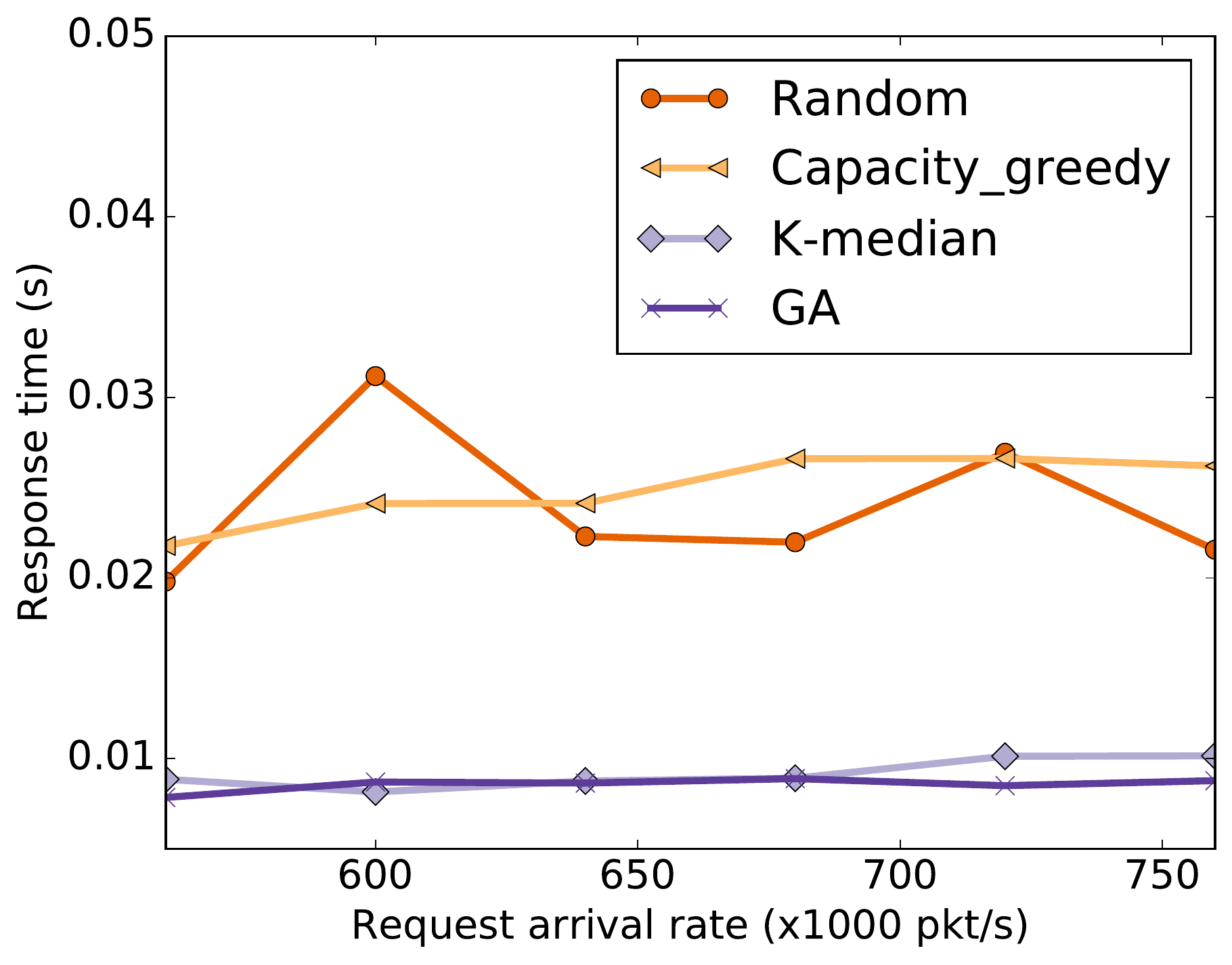}}
    \subfloat[Utilization]{\label{scen3_util}\includegraphics[width=.5\linewidth]{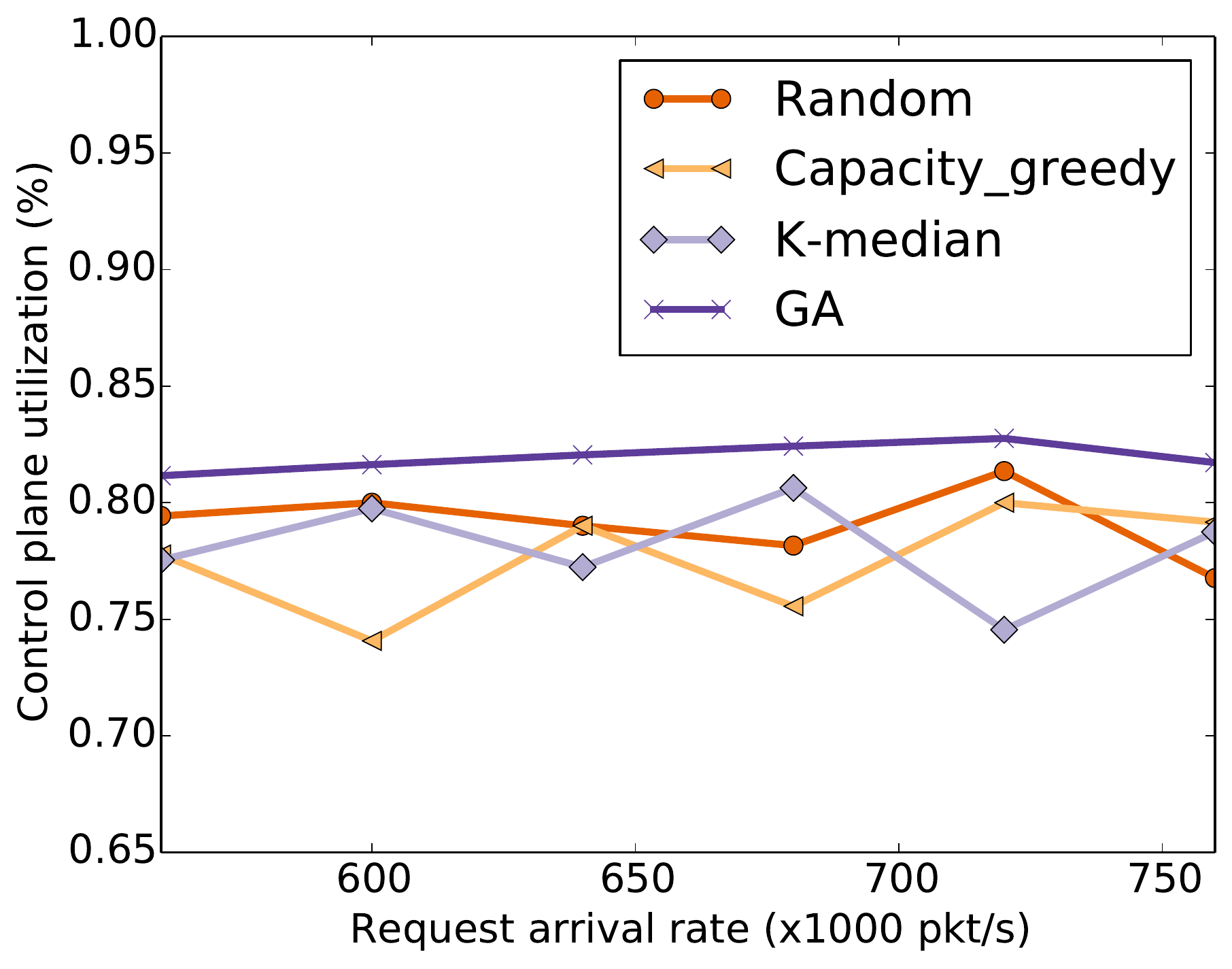}}
  \end{center}
  \caption{The performance comparison of different methods for controller placement in a large scale network with heterogeneous controller setting.}
\label{fig:scen3_placement_compa}
\vspace{-15pt}
\end{figure}

\begin{figure}[!tbp]
  \begin{center}
    \subfloat[small-scale network]{\label{scen2_ga}\includegraphics[width=.5\linewidth]{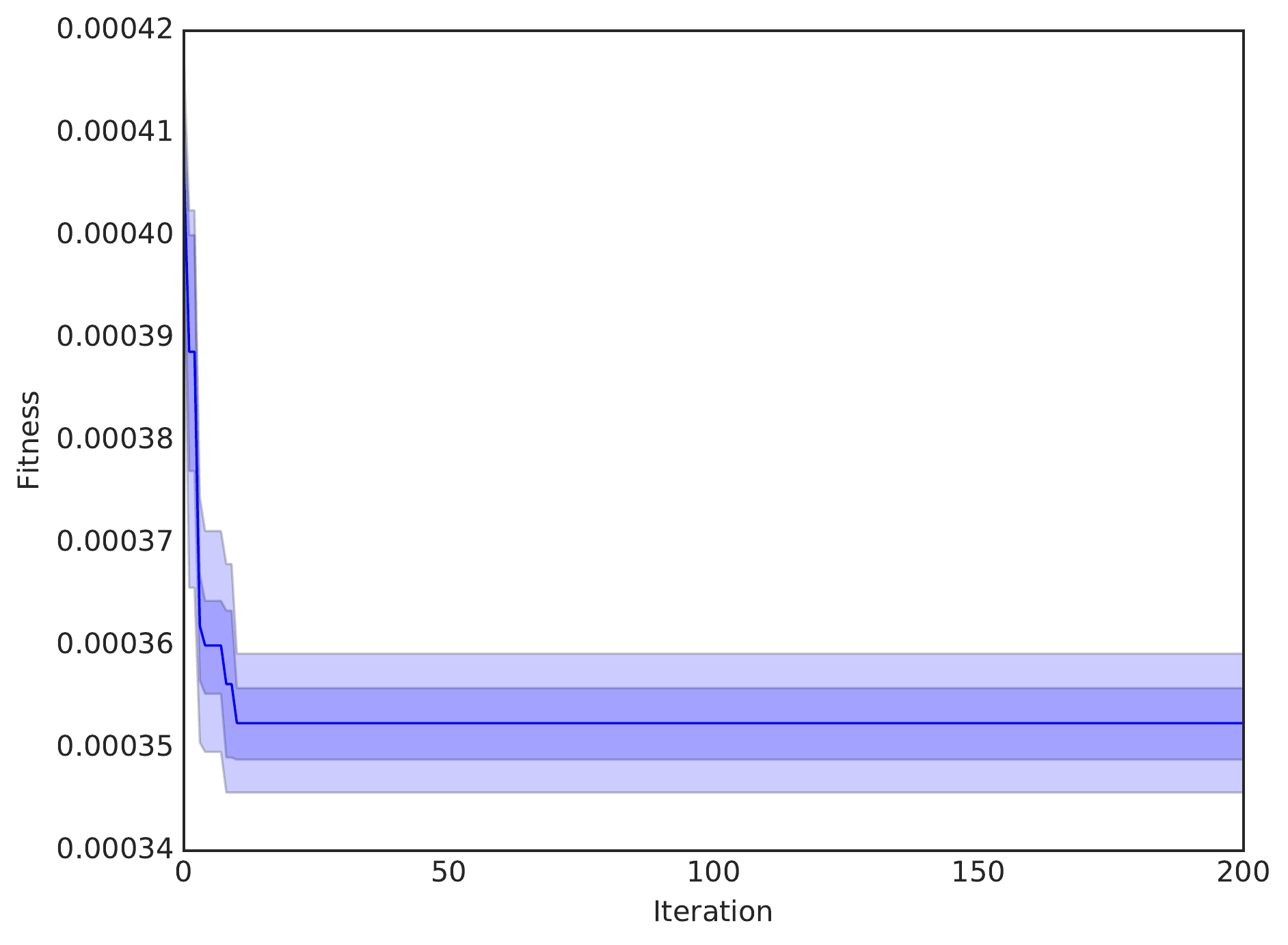}}
    \subfloat[large-scale network]{\label{scen3_ga}\includegraphics[width=.5\linewidth]{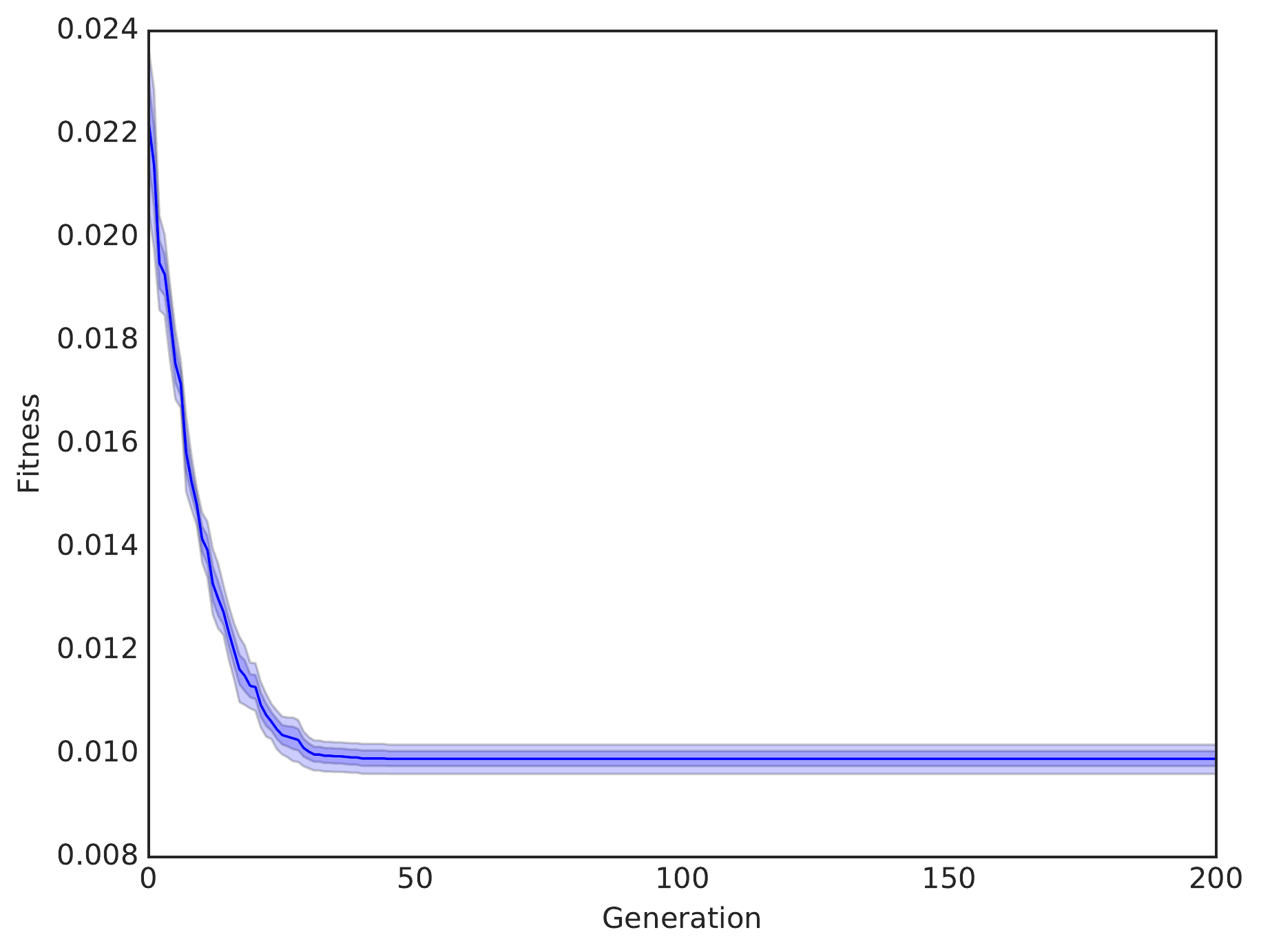}}
  \end{center}
  \caption{The convergence of GA for controller placement problem in a network with heterogeneous controller setting.}
\label{fig:scen3_ga}
\vspace{-10pt}
\end{figure}

To further demonstrate the effectiveness of GA, we run different placement methods at various request arrival rates. Fig. \ref{fig:scen3_placement_compa} depicts the performance comparison in terms of the packet response time and control plane utilization. It can be seen from Fig. \ref{fig:scen3_placement_compa}(a) that the response time obtained by using both GA and K-median remain steadily below $10$ ms. In certain situations, GA even performs better. In comparison, the response time of the random approach shows large fluctuation ranging from $20$ ms to $32$ ms due to the randomly chosen remote controllers. In terms of the control plane utilization, we notice that GA achieves the highest utilization which is around $82\%$ (closely matches the decay factor $\boldsymbol \beta = 0.83$ we set in Section \ref{subsec:exp_setup} ) and remains nearly constant regardless the traffic changes. 
In comparison, the utilization of other methods varies dramatically with different packet arrival rates. For example, the lowest utilization which both capacity-based greedy and K-median methods reach is $74\%$ while their highest utilization can be $80\%$. This is mainly because these methods choose controllers solely based on their capacities or communication delay, without carefully considering or even completely ignoring the control plane utilization. Thus, the utilization of the control plane can vary dramatically as the request arrival rate changes.

We also investigate the convergence speed of GA by running the algorithm for $30$ times with two different network settings (i.e., Section \ref{subsec:small-dc-diff-ctl} and Section \ref{subsec:large-network-diff-ctl}) respectively. In each run, GA is set with $200$ generations with a population size $50$ for the small-scale network and $200$ for the large-scale network. Fig. \ref{fig:scen3_ga} depicts the change of fitness value across generations with both $68\%$ and $95\%$ confidence bands. From Fig. \ref{fig:scen3_ga}, we find that GA converges quickly regardless of the problem size. In the small-scale network, GA can converge within $10$ generations. In the large network, the algorithm can still converge within $30$ generations. 

\section{Conclusions} \label{sec:conclusion}
When deploying distributed controllers in an SDN network, one essential issue is the controller placement problem (CPP). Existing works mainly focus on reducing the communication delay while undermining the significance of the workload distribution among controllers. 
However, previous works show that the performance of the CPP can be highly related to the workload distribution. 
In this paper, a queuing model is built which systematically measures the relationship among the network response time, the workload distribution and the communication delay. 
Meanwhile, to maintain reasonable operation cost, the control plane utilization must be kept at a high level. 
Motivated by this, we formulate the CPP as a constrained optimization problem using the built queuing model, which simultaneously optimizes the response time and the control plane utilization. Several alternative ways are investigated to solve the CPP, ranging from simple heuristic approaches to more advanced search methods. In particular, a new algorithm combining the use of GA and GD is proposed. The performance of our algorithm is analyzed in detail via a series of simulation featuring different network settings. It is shown that our algorithm achieved highest control plane utilization and competitively low response time compared to the widely-used heuristic methods.

\bibliographystyle{IEEEtranS}
\bibliography{im_v4}
\end{document}